# ZAGNet: Zone-Aware Graph Aggregation Network for Patient-Level Lung Ultrasound Diagnosis

Li Chen, Shubham Patil, Rashid Al Mukaddim, Jochen Kruecker, Balasundar Raju and Alvin Chen
*Philips Ultrasound*
Cambridge, MA, 02141
Email: {li.chen_1, shubham.patil, rashidal.mukaddim, jochen.kruecker, balasundar.raju, alvin.chen}@philips.com

***Abstract*— Patient-level lung ultrasound (LUS) diagnosis requires integrating findings acquired across multiple anatomical zones, yet clinical examinations frequently involve variable and incomplete scanning protocols with missing zones. Existing diagnostic AI methods primarily analyze individual frames or video loops, relying on heuristic aggregation strategies such as max or mean pooling that ignore inter-zone relationships for patient-level inference. This paper presents ZAGNet, a Zone-Aware Graph Neural Network that represents temporally tracked pathology findings as graph nodes connected by anatomical zone adjacency. A graph transformer network propagates contextual information across neighboring lung regions, while a virtual global node aggregates graph-level features to predict patient-level consolidation and pleural effusion using only patient-level supervision. ZAGNet accommodates missing zones by computing on a graph structure without fixed input format or size. We evaluate ZAGNet on a multicenter dataset of 714 subjects (20,256 LUS video loops) with exams varying from 4 to 16 zones across anterior, posterior, and lateral thoracic regions. For consolidation diagnosis, ZAGNet achieved an AUC of 0.803 compared to 0.677 (max pooling) and 0.674 (mean pooling). For pleural effusion, AUC increased to 0.893 from 0.804 (max pooling) and 0.815 (mean pooling). These represent improvements of up to 19% and 11% for consolidation and pleural effusion respectively. The results demonstrate that graph-based inter-zone reasoning provides an effective and clinically consistent framework for automated patient-level LUS assessment.**

## I. Introduction

Point-of-care lung ultrasound (LUS) is widely used in emergency and critical care settings for bedside evaluation of pulmonary conditions including consolidation and pleural effusion [1], [2]. During an examination, clinicians sequentially acquire video loops from multiple predefined anatomical zones and integrate regional findings to reach a patient-level diagnosis. The final clinical assessment therefore depends not only on the presence of pathology within each zone, but also on its spatial distribution across the lung.

Recent advances in deep learning have substantially improved automated LUS analysis, including frame-level pathology detection, video classification, and segmentation of lung artifacts [3]–[7]. However, most existing methods focus on analyzing individual frames or video loops rather than directly predicting patient-level diagnoses. Patient-level inference is consequently obtained by applying heuristic aggregation strategies such as max or mean pooling to regional predictions. These approaches ignore anatomical relationships between lung zones and cannot model how multiple regional findings jointly contribute to a patient-level diagnosis. Moreover, clinical LUS examinations frequently follow abbreviated or opportunistic scanning protocols, resulting in missing or variable numbers of scanned zones – a variability that conventional aggregation methods are not designed to handle.

Graph neural networks (GNNs) provide a natural framework for modeling structured relationships among heterogeneous observations through message passing over graph topology [8], [9]. Graph-based reasoning has recently demonstrated promising results in medical image analysis by aggregating information from multiple image regions into patient-level predictions [10]. ThyGraph, for example, formulates thyroid ultrasound diagnosis as graph reasoning over local image patches and demonstrates the effectiveness of graph-based aggregation compared with conventional pooling strategies [11]. However, existing approaches do not address the unique acquisition characteristics of LUS, where clinical examinations consist of variable numbers of anatomically organized scanning zones and multiple pathological findings may coexist within the same zone.

This paper proposes ZAGNet, a Zone-Aware Graph Neural Network for patient-level LUS diagnosis, with the following contributions. First, rather than representing each scan zone by a single feature vector, ZAGNet constructs a patient-level graph whose nodes correspond to temporally tracked pathology findings; this allows multiple independent findings within the same zone to be modeled separately. Second, nodes are connected according to a predefined anatomical adjacency map, enabling contextual information to propagate across neighboring lung regions. Third, a virtual global node aggregates graph-level representations to produce patient-level predictions under weak supervision using only patient-level labels. Because the graph is constructed from detected findings rather than a fixed zone grid, ZAGNet naturally accommodates variable and incomplete scanning protocols without padding or imputation. Experiments on a multicenter LUS dataset demonstrate that ZAGNet substantially outperforms

This work was supported in part by the U.S. Department of Health and Human Services (HHS); Administration for Strategic Preparedness and Response (ASPR); Biomedical Advanced Research and Development Authority (BARDA), under contract number 75A50120C00097.

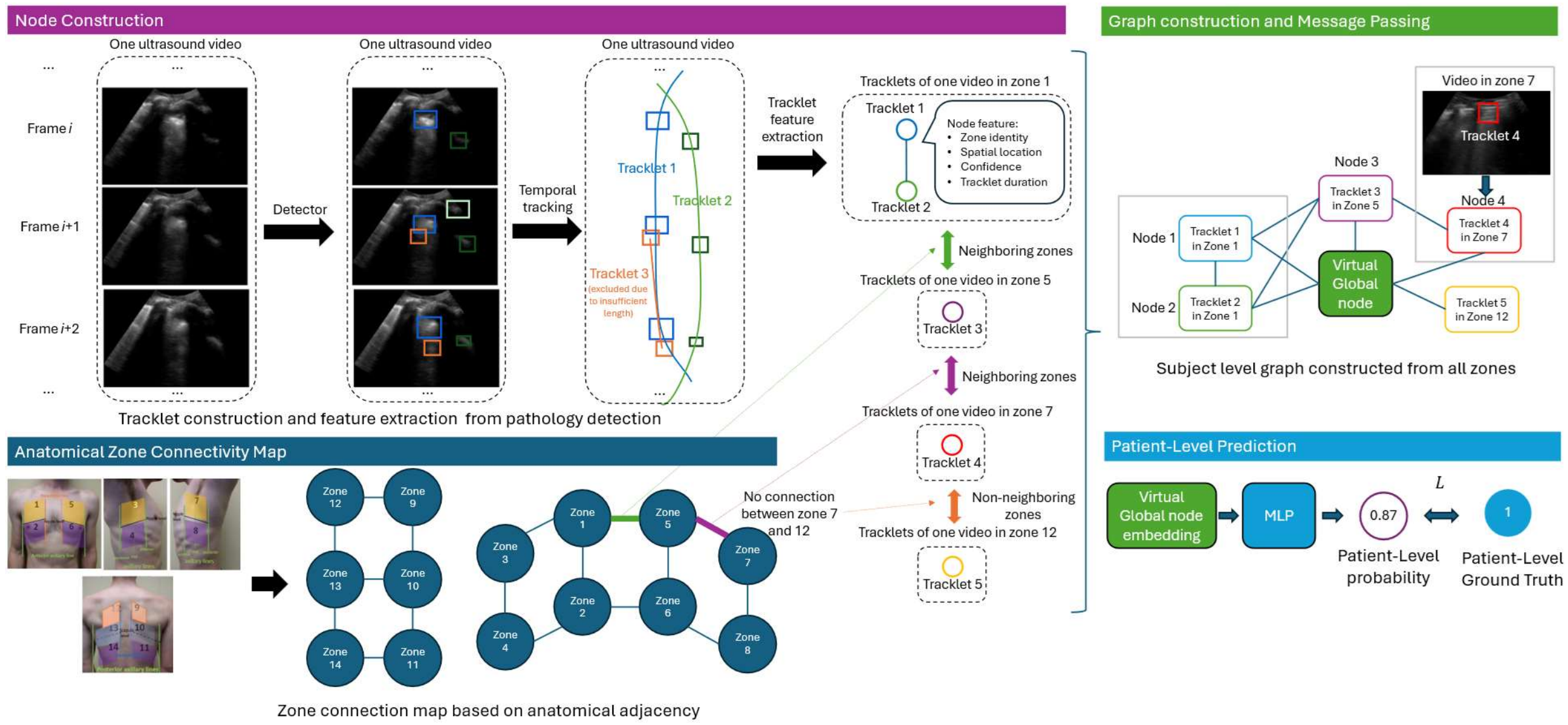

Fig. 1 Overview of the proposed ZAGNet framework, consisting of three stages: (1) node construction from temporally tracked pathology findings, (2) graph construction based on anatomical zone adjacency, and (3) graph transformer reasoning for patient-level prediction.

conventional pooling-based aggregation for both consolidation and pleural effusion classification.

## II. Methods

### A. Problem Formulation

We formulate patient-level LUS diagnosis as a graph-level classification problem. Given a set of LUS video loops $\{v_1, v_2, ..., v_N\}$ acquired across N anatomical zones for a patient, the goal is to predict a binary patient-level label $y \in \{0,1\}$ for each pathology of interest (consolidation and pleural effusion). Rather than aggregating zone-level predictions directly, ZAGNet converts variable-length LUS examinations into patient-specific graph representations $G = (V, E)$ constructed from detected pathological findings. As illustrated in Fig. 1, the framework consists of three stages: (1) node construction from temporally tracked pathological findings, (2) graph construction based on anatomical relationships between lung zones, and (3) graph transformer reasoning for patient-level prediction.

### B. Node Construction via Temporal Tracking

ZAGNet represents each temporally consistent pathology tracklet as a graph node rather than operating directly on frame-level detections. Frame-level pathology detections are first generated from each LUS video using a pretrained object detector. Consecutive detections belonging to the same pathological finding are associated into tracklets using the Simple Online and Realtime Tracking (SORT) algorithm [12]. Tracklets shorter than five frames are discarded to suppress unstable detections and reduce false positives. This design exploits the temporal redundancy naturally present in ultrasound video loops while substantially reducing graph complexity; a pathological finding typically appears across many consecutive frames, and representing these observations with a single node removes redundancy, stabilizes detection confidence through temporal association, and enables efficient graph reasoning. Importantly, representing tracklets rather than scan zones allows multiple independent findings within the same anatomical zone to be modeled separately.

Each node $v_i \in V$ corresponds to one tracklet and is represented by an 8-dimensional feature vector:

$$\boldsymbol{x}_i = [z_i, x_i, y_i, w_i, h_i, c_i, d_i, p_i], \tag{1}$$

where $z_i$ is a one-hot encoded zone identifier, $(x_i, y_i, w_i, h_i)$ are normalized bounding box coordinates, $c_i$ is the mean detection confidence, $d_i$ is the normalized tracklet duration, and $p_i$ is a one-hot encoded pathology class label. These features jointly encode the anatomical location, spatial extent, temporal persistence, and semantic information of each finding.

### C. Graph Construction

Edges are defined by two connectivity rules reflecting the anatomical organization of standard LUS acquisition protocols:

1. **Within-zone edges** connect all tracklet nodes within the same anatomical zone, capturing intra-zone co-occurrence of findings.
2. **Inter-zone edges** connect nodes from anatomically adjacent zones according to a predefined zone adjacency map derived from the clinical scanning protocol, in which up to 16 zones are distributed across the anterior, posterior, and lateral chest wall.

Unlike fully connected graphs, this design incorporates prior anatomical knowledge while avoiding message passing between unrelated lung regions. Because nodes are generated only from observed tracklets, missing scan zones naturally result in absent graph nodes without requiring padding or imputation.

To obtain a graph-level representation, a virtual global node $v_g$ is added and connected to every tracklet node. Following the virtual-node design commonly adopted in graph representation learning [10], the global node exchanges information with all pathology nodes during message passing and serves as the aggregated representation of the entire examination — providing a richer, context-aware summary than mean or max pooling over individual node embeddings.

### D. Graph Reasoning

Node features are first projected into a shared embedding space using a linear layer. Graph reasoning is then performed using a graph transformer convolution layer (TransformerConv) [13], which computes attention-based message passing between neighboring nodes. At the $l$-th layer, the hidden representation of node $i$is updated by aggregating features from its neighbors:

$$h_i^{(l+1)} = \sigma(\sum_{j \in N(i)} \alpha_{ij} \mathbf{W} h_j^{(l)}), \tag{2}$$

where $\mathcal{N}(i)$ denotes the neighbors of node $i$ defined by the graph topology, $\alpha_{ij}$ are learned attention coefficients, $\mathbf{W}$ is a learnable linear transformation, and $\sigma(\cdot)$ is a nonlinear activation function. Through iterative message passing, each node progressively incorporates contextual information from anatomically related pathological findings.

### E. Patient-Level Prediction

After the final graph transformer layer, the embedding of the virtual global node $\mathbf{h}_g$ summarizes pathological information across the entire examination. The patient-level prediction is obtained via a linear classification head:

$$\hat{y} = \sigma(\mathbf{W}_c \mathbf{h}_g + \mathbf{b}), \tag{3}$$

where $\mathbf{W}_c$ and $\mathbf{b}$ are learnable parameters and $\sigma(\cdot)$ is the sigmoid activation. The network is trained using only patient-level labels without requiring zone-level supervision. For each binary classification task (consolidation or pleural effusion), the objective is the binary cross-entropy loss,

$$L = -y log(\hat{y}) - (1 - y) log(1 - \hat{y}), \tag{4}$$

where $y \in \{0,1\}$ is the ground-truth patient label. By operating directly on graph representations, ZAGNet naturally accommodates examinations with variable numbers of findings and incomplete scanning protocols, making it well suited for real-world point-of-care LUS applications.

## III. Experiments and Results

### A. Dataset

The proposed method was evaluated on a multicenter LUS dataset of 714 subjects (20,256 video loops) from emergency and ICU settings across 11 clinical sites. Each examination contained 4–16 anatomical scan zones covering anterior, lateral, and posterior thoracic regions. Patient-level ground truth for consolidation and pleural effusion was established by clinicians using all available clinical information, including LUS findings, CT imaging, laboratory tests, and medical history. The dataset was divided into training, validation, and testing sets using subject-level partitioning.

### B. Implementation Details

ZAGNet was implemented using PyTorch. The model contains 34,265 trainable parameters and requires 13.4K FLOPs per inference. The model was trained using the Adam optimizer with a learning rate of 1e-4 and 500 epochs. Only patient-level labels were used for model training without any zone-level supervision.

### C. Experiments

Two heuristic aggregation strategies were implemented as baselines by replacing the ZAGNet graph reasoning module with max pooling or mean pooling over node features. This ensures that performance differences are attributable to the aggregation strategy rather than the underlying detection model. Model performance was evaluated using the area under the receiver operating characteristic curve (AUC).

### D. Results

ZAGNet consistently outperformed both pooling-based aggregation methods for patient-level diagnosis (TABLE I). For consolidation, AUC increased from 0.677 (max pooling) and 0.674 (mean pooling) to 0.803, representing improvements of up to 19%. For pleural effusion, AUC increased from 0.804 (max pooling) and 0.815 (mean pooling) to 0.893, representing improvements of up to 11%. These results demonstrate that explicitly modeling anatomical zone relationships provides more informative patient-level representations than independent regional aggregation. By propagating contextual information between neighboring regions through graph message passing, ZAGNet enables more clinically consistent integration of findings while naturally accommodating variable and incomplete scanning protocols.

TABLE I. Comparison of LUS zone aggregation methods for patient-level diagnostic classification (AUC).

| Method | Consolidation | Pleural Effusion |
|---|---|---|
| Max Pooling | 0.677 | 0.804 |
| Mean Pooling | 0.674 | 0.815 |
| **ZAGNet (proposed)** | **0.803** | **0.893** |

## IV. Conclusion

This paper presented ZAGNet, a graph-based framework for patient-level lung ultrasound diagnosis. By representing tracked pathological findings as graph nodes connected through anatomical zone relationships, ZAGNet enables structured reasoning across scan regions while naturally handling variable and incomplete exams using only patient-level information. Experiments on a multicenter dataset demonstrated substantial improvements over conventional pooling-based aggregation for consolidation and pleural effusion classification. Future work will investigate learned graph topologies, joint training of detection and aggregation modules, additional pulmonary pathologies, and validation on more diverse clinical datasets.

## Acknowledgment

We gratefully acknowledge C. L. Moore, C. Baloescu, R. Denkewicz, M. Hesami, B. Hicks, N. Schnittke, M. Parker, A. Silver, M. Petrovich, A. Sabbaj, L. Hu, D. Chan, M. Zhu, M. Kaili, C. Thomas, M. Feuerherdt, Y. Zhang, C.R. Gregory, K.W. Gregory, D.O. Kessler, D. Coneybeare, L. Malia, J. Shupp, C. Mehanian, S. Kulhare, and R. Millin for their contributions to clinical data acquisition, curation, and annotation. This work was supported in part by the U.S. Department of Health and Human Services; Administration for Strategic Preparedness and Response; Biomedical Advanced Research and Development Authority, under contract 75A50120C00097. The findings do not necessarily represent the views of HHS or its components.